\documentclass[prb,twocolumn,showpacs]{revtex4}%
\usepackage{amsfonts}
\usepackage{amsmath}
\usepackage{amssymb}
\usepackage{graphicx}%
\providecommand{\U}[1]{\protect\rule{.1in}{.1in}}
\begin{document}
\preprint{HEP/123-qed}
\title[ ]{Theory of coexistence of superconductivity and ferroelectricity}

\begin{abstract}
A new investigation of the coexistence and competition of ferroelectricity and
superconductivity is reported. In particular we show that the starting
Hamiltonian of a previous study by Birman and Weger (2001) can be exactly
diagonalized. The result differs significantly from mean-field theory. A
Hamiltonian with a different realization of the coupling between
ferroelectricity and superconductivity is proposed. We report the results for
mean-field theory applied to this Hamiltonian. We find that the order
parameters are strongly affected by this coupling.

\end{abstract}
\author{Y. Krivolapov}
\email{evgkr@tx.technion.ac.il}
\author{A. Mann}
\author{Joseph L. Birman}
\altaffiliation[Permanent address: ]{Department of Physics, City College, CUNY, New York, New York 10031}

\affiliation{Department of Physics, Technion-Israel Institute of Technology, Haifa 32000 Israel}
\volumeyear{year}
\volumenumber{number}
\issuenumber{number}
\eid{identifier}
\date{\today}
\received[Received text]{date}

\revised[Revised text]{date}

\accepted[Accepted text]{date}

\published[Published text]{date}

\startpage{1}
\endpage{2}
\maketitle

In the present paper we report two related results from a reexamination of
previous work on nearly ferroelectric superconductors \cite{Birman}. In that
paper a coupling term was introduced into the original Hamiltonian (Eq. 27 of
Ref.~\onlinecite{Birman}). An investigation of that coupling term shows that
it only gives a squeezing of the phonons and \emph{no }coupling to the
electronic pairs, and therefore the Hamiltonian can be diagonalized exactly.
However, the double mean-field approximation does result in an effective
coupling between the two subsystems. (Eq. 34 of Ref.~\onlinecite{Birman}).
Therefore the results of the analysis of that equation remain valid. Our
second result follows from introducing a different biquadratic coupling term
in the Hamiltonian of Eq.~27 of Ref.~\onlinecite{Birman}, which satisfies
gauge and inversion symmetries. This term does couple the two subsystems. We
will treat this new Hamiltonian in mean-field approximation.

We start with the model of coexistence of superconductivity and
ferroelectricity proposed by Birman and Weger \cite{Birman}. For convenience
we include its main features here. Birman and Weger start with two separate
Hamiltonians for the superconducting and ferroelectric sectors. The
superconducting sector is a mean-field reduced BCS pseudo-spin model (we
corrected a misprint in Ref.~\onlinecite{Birman})%
\[
H_{SC}=-2\sum_{k}\left(  \varepsilon_{k}\hat{\jmath}_{3k}+2\Delta_{k}%
\hat{\jmath}_{2k}\right)
\]
where $\varepsilon_{k}$ is the single-electron energy and $\Delta_{k}$ is the
pairing interaction energy. The pseudo-spin operators $\hat{\jmath}_{pk}$ obey
$SU\left(  2\right)  $ commutation relations and are defined as%
\begin{align*}
\hat{\jmath}_{1k} &  =\left(  -i/2\right)  \left(  \hat{b}_{k}^{\dag}-\hat
{b}_{k}\right)  \\
\hat{\jmath}_{2k} &  =\left(  1/2\right)  \left(  \hat{b}_{k}^{\dag}+\hat
{b}_{k}\right)  \\
\hat{\jmath}_{3k} &  =\left(  -1/2\right)  \left(  \hat{n}_{k}+\hat{n}%
_{-k}-1\right)
\end{align*}
where \ $\hat{b}_{k}$ and $\hat{b}_{k}^{\dag}$ are pair operators defined by
\[
\hat{b}_{k}^{\dag}=\hat{a}_{k\uparrow}^{\dag}\hat{a}_{-k\downarrow}^{\dag
},\text{ \ \ }\hat{b}_{k}=\left(  \hat{b}_{k}^{\dag}\right)  ^{\dag},\text{
\ \ }\hat{n}_{k}\equiv\hat{a}_{k\uparrow}^{\dag}\hat{a}_{k\uparrow}%
\]
and $\hat{a}_{k\uparrow}$ and $\hat{a}_{k\uparrow}^{\dagger}$ are the electron
annihilation and creation operators for wave vector $\vec{k},$ spin $\left(
\uparrow\right)  ,$ etc.

The Hamiltonian of the ferroelectric sector of the model is simply a
Hamiltonian of a displaced harmonic oscillator%
\[
H_{FE}=\omega_{0}\left(  \hat{N}_{B}+\frac{1}{2}\right)  +\gamma_{1}\left(
\hat{B}_{0}^{\dag}+\hat{B}_{0}\right)
\]
where $\hat{B}_{0}^{\dag},\hat{B}_{0}$ have boson commutation relations.

Combining those two sectors together and introducing a gauge invariant
coupling between them, Birman and Weger obtain the following Hamiltonian
(Eq.~27 of Ref.~\onlinecite{Birman})%

\begin{align}
\hat{H}  &  =-2\sum_{k}\left(  \varepsilon_{k}\hat{\jmath}_{3k}+2\Delta
_{k}\hat{\jmath}_{2k}\right)  +\omega_{0}\left(  \hat{N}_{B}+\frac{1}%
{2}\right)  +\\
&  +\gamma_{1}\left(  \hat{B}_{0}^{\dag}+\hat{B}_{0}\right)  +\sum_{k}%
\gamma_{2k}\hat{\jmath}_{2k}^{2}\left(  \hat{B}_{0}^{\dag}+\hat{B}_{0}\right)
^{2}\nonumber
\end{align}

If a single $k$ mode is isolated%
\begin{align}
\hat{H}  &  =-2\left(  \varepsilon\hat{\jmath}_{3}+2\Delta\hat{\jmath}%
_{2}\right)  +\omega_{0}\left(  \hat{N}_{B}+\frac{1}{2}\right)  +\\
&  +\gamma_{1}\left(  \hat{B}_{0}^{\dag}+\hat{B}_{0}\right)  +\gamma_{2}%
\hat{\jmath}_{2}^{2}\left(  \hat{B}_{0}^{\dag}+\hat{B}_{0}\right)
^{2}\nonumber
\end{align}

Here we notice that the structure of the pseudo-spin operators in terms of the
pair operators implies%
\[
\hat{\jmath}_{1}^{2}=\hat{\jmath}_{2}^{2}=\hat{\jmath}_{3}^{2}%
\]
and therefore $\hat{\jmath}_{1}^{2},$ $\hat{\jmath}_{2}^{2}$ and $\hat{\jmath
}_{3}^{2}$ are invariant to any rotation of the form $\exp\left[  \hat{n}%
\cdot\overrightarrow{j}\right]  .$

We use this property to diagonalize the Hamiltonian analytically by applying
the following unitary transformation%
\begin{align*}
&  e^{\frac{\beta}{2}\left(  \hat{B}_{0}^{\dag2}-\hat{B}_{0}^{2}\right)
}e^{\alpha\left(  \hat{B}_{0}^{\dag}-\hat{B}_{0}\right)  +i\theta\hat{\jmath
}_{1}}He^{-i\theta\hat{\jmath}_{1}-\alpha\left(  \hat{B}_{0}^{\dag}-\hat
{B}_{0}\right)  }e^{-\frac{\beta}{2}\left(  \hat{B}_{0}^{\dag2}-\hat{B}%
_{0}^{2}\right)  }\\
&  =H_{D}+\hat{\jmath}_{2}\left(  2\varepsilon\sin\theta-4\Delta\cos
\theta\right)  +\\
&  +\left(  \hat{B}_{0}^{\dag}+\hat{B}_{0}\right)  \left(  \omega_{0}\alpha
e^{-2\beta}+\gamma_{1}e^{-\beta}+4\alpha\gamma_{2}\hat{\jmath}_{2}%
^{2}e^{-2\beta}\right)  +\\
&  +\left(  \hat{B}_{0}^{\dag2}+\hat{B}_{0}^{2}\right)  \left(  -\frac
{\omega_{0}}{2}\operatorname*{sh}2\beta+\gamma_{2}\hat{\jmath}_{2}%
^{2}e^{-2\beta}\right)
\end{align*}
Here $H_{D}$ is a Hamiltonian diagonal in the $\left\vert j_{3},N_{B}%
\right\rangle $ basis:%
\begin{align*}
H_{D} &  =2\hat{\jmath}_{3}\left(  2\Delta\sin\theta-\varepsilon\cos
\theta\right)  +\\
&  +\omega_{0}\left(  \hat{N}_{B}\operatorname*{ch}2\beta+\operatorname*{sh}%
\nolimits^{2}\beta+\frac{1}{2}+\alpha e^{-2\beta}\right)  +\\
&  +2\alpha\gamma_{1}e^{-\beta}+\gamma_{2}\hat{\jmath}_{2}^{2}e^{-2\beta
}\left(  2\hat{N}_{B}+1+4\alpha^{2}\right)
\end{align*}
and if we want the non-diagonal parts to vanish we require%
\begin{align*}
2\varepsilon\sin\theta &  =4\Delta\cos\theta\\
\omega_{0}\alpha e^{-2\beta} &  +\gamma_{1}e^{-\beta}+4\alpha\gamma_{2}%
\hat{\jmath}_{2}^{2}e^{-2\beta}=0\\
\gamma_{2}\hat{\jmath}_{2}^{2}e^{-2\beta} &  =\frac{\omega_{0}}{2}%
\operatorname*{sh}2\beta
\end{align*}
These yield the following set of coupled equations for the parameters%
\[
\tan\theta=\frac{2\Delta}{\varepsilon},\qquad\alpha=-\frac{\gamma_{1}e^{\beta
}}{\omega_{0}+\gamma_{2}},\qquad e^{4\beta}=\frac{\omega_{0}+\gamma_{2}%
}{\omega_{0}}.
\]
Inserting them back into the transformed Hamiltonian gives%
\begin{align*}
H_{D} &  =-2\hat{\jmath}_{3}\frac{\varepsilon^{2}-4\Delta^{2}}{\sqrt
{\varepsilon^{2}+4\Delta^{2}}}-\frac{\gamma_{1}^{2}}{\gamma_{2}+\omega_{0}}+\\
&  +\sqrt{\omega_{0}\left(  \omega_{0}+\gamma_{2}\right)  }\left(  \hat{N}%
_{B}+\frac{1}{2}\right)
\end{align*}
and this is the decoupled Hamiltonian of a squeezed and shifted harmonic
oscillator and a spin of one-half in a magnetic field. The energy spectrum is
given by%
\begin{align*}
H_{D}\left(  m,n\right)   &  =-2m\frac{\varepsilon^{2}-4\Delta^{2}}%
{\sqrt{\varepsilon^{2}+4\Delta^{2}}}-\frac{\gamma_{1}^{2}}{\gamma_{2}%
+\omega_{0}}+\\
&  +\sqrt{\omega_{0}\left(  \omega_{0}+\gamma_{2}\right)  }\left(  n+\frac
{1}{2}\right)
\end{align*}

We can now use the exact eigenstates of the Hamiltonian to compute the order
parameters of the system:%
\begin{align}
\eta_{FE} &  \equiv\left\langle \hat{B}_{0}^{\dag}+\hat{B}_{0}\right\rangle
=2\alpha e^{-\beta}=-\frac{\gamma_{1}}{\omega_{0}+\gamma_{2}}\label{app:Exact}%
\\
\eta_{SC} &  \equiv\left\langle \hat{\jmath}_{2}\right\rangle =-\frac{1}%
{2}\sin\theta=-\frac{\Delta}{\sqrt{\varepsilon^{2}+4\Delta^{2}}}\nonumber
\end{align}
While the order parameter of the ferroelectric sector is affected by the SC-FE
coupling strength, the order parameter of the superconducting sector is not.
This occurs since in the coupling term $\gamma_{2}\hat{\jmath}_{2}^{2}\left(
\hat{B}_{0}^{\dag}+\hat{B}_{0}\right)  ^{2},$ $\hat{\jmath}_{2}^{2}$ is a
c-number and not an operator.

Using a double mean-field approximation for the Hamiltonian (Eq.~34 of
Ref.~\onlinecite{Birman}) and variational coherent state wave function, Birman
and Weger \cite{Birman} obtained the following relations for the order
parameters%
\[
\eta_{FE}=2\xi\qquad\eta_{SC}=\frac{m\left(  \Delta^{\prime}/\varepsilon
+\xi\Gamma_{2}/\varepsilon\right)  }{\sqrt{1+\left(  \Delta^{\prime
}/\varepsilon+\xi\Gamma_{2}/\varepsilon\right)  ^{2}}}%
\]
with%
\begin{align*}
\xi &  =\left[  -\Gamma_{1}/\omega_{0}-\Gamma_{2}\left(  1-m\sin\theta\right)
\right]  /\omega_{0}\\
\Delta^{\prime}  &  =\Delta\left(  1-\gamma_{2}P^{2}\right) \\
\Gamma_{1}  &  =\gamma_{1}-2\gamma_{2}P\Delta^{2}\qquad\Gamma_{2}=4\gamma_{2}%
\end{align*}
Here both order parameters are affected by the SC-FE coupling strength
$\left(  \gamma_{2}\right)  .$

To sum up this part, we have obtained an exact analytical solution of the
ferroelectricity and superconductivity coexistence model, used in
Ref.~\onlinecite{Birman}. We showed here that contrary to the mean-field
results, an exact solution of the model demonstrates that the ferroelectric
order parameter is affected by the coupling, but the superconducting order
parameter is not. If one were to initially model the SC-FE system by Eq.~34 of
Ref.~\onlinecite{Birman}, the results reported in Ref.~\onlinecite{Birman}
would remain valid.

Returning to the original question of the proper coupling term in a quantum
mechanical Hamiltonian, we need to respect gauge and inversion symmetry. Since
the superconducting gap parameter is a complex quantity the added term should
correspond to the bilinear $\left\vert \Delta\right\vert ^{2}.$ We now take
this as $\hat{\jmath}_{+}\hat{\jmath}_{-}.$ The coupling term for the
ferroelectric polarization, which respects inversion symmetry will correspond
to $P^{2}$ and can be taken as $\left(  \hat{B}_{0}^{\dag}+\hat{B}_{0}\right)
^{2}.$ Thus the bilinear coupling becomes $\hat{\jmath}_{+}\hat{\jmath}%
_{-}\left(  \hat{B}_{0}^{\dag}+\hat{B}_{0}\right)  ^{2}.$ We note that
$\hat{\jmath}_{+}\hat{\jmath}_{-}=\hat{\jmath}_{1}^{2}+\hat{\jmath}_{2}%
^{2}+j_{3}=\frac{1}{2}+\hat{\jmath}_{3}$ (Using $\hat{\jmath}_{-}\hat{\jmath
}_{+}$ would have changed $\hat{\jmath}_{3}$ into $-\hat{\jmath}_{3}$). The
coupling becomes $\gamma_{2}\left(  \hat{\jmath}_{3}+\frac{1}{2}\right)
\left(  \hat{B}_{0}^{\dag}+\hat{B}_{0}\right)  ^{2}.$ The Hamiltonian is now%

\begin{align}
\hat{H} &  =-2\left(  \varepsilon\hat{\jmath}_{3}+2\Delta\hat{\jmath}%
_{2}\right)  +\omega_{0}\left(  \hat{N}_{B}+\frac{1}{2}\right)
+\label{eq:newH}\\
&  +\gamma_{1}\left(  \hat{B}_{0}^{\dag}+\hat{B}_{0}\right)  +\gamma
_{2}\left(  \hat{\jmath}_{3}+\frac{1}{2}\right)  \left(  \hat{B}_{0}^{\dag
}+\hat{B}_{0}\right)  ^{2}\nonumber
\end{align}
The interaction term in this Hamiltonian differs from the corresponding term
(Eq.~27) in Ref.~\onlinecite{Birman}. This Hamiltonian is not exactly
solvable, therefore we make a mean-field approximation in the form%
\begin{align}
\hat{\jmath}_{3}\left(  \hat{B}_{0}^{\dag}+\hat{B}_{0}\right)  ^{2} &
\simeq\left\langle \left(  \hat{B}_{0}^{\dag}+\hat{B}_{0}\right)
^{2}\right\rangle \hat{\jmath}_{3}+\label{eq:newMF}\\
&  +\left\langle \hat{\jmath}_{3}\right\rangle \left(  \hat{B}_{0}^{\dag}%
+\hat{B}_{0}\right)  ^{2}-\left\langle \left(  \hat{B}_{0}^{\dag}+\hat{B}%
_{0}\right)  ^{2}\right\rangle \left\langle \hat{\jmath}_{3}\right\rangle
.\nonumber
\end{align}
Inserting Eq.~\ref{eq:newMF} into Eq.~\ref{eq:newH} results in a solvable
bilinear Hamiltonian%
\begin{align}
\hat{H}_{MF} &  =\varepsilon^{\prime}\hat{\jmath}_{3}-4\Delta\hat{\jmath}%
_{2}+\omega_{0}\left(  \hat{N}_{B}+\frac{1}{2}\right)  +\gamma_{1}\left(
\hat{B}_{0}^{\dag}+\hat{B}_{0}\right)  +\label{eq:MF}\\
&  +\gamma_{2}^{\prime}\left(  \hat{B}_{0}^{\dag}+\hat{B}_{0}\right)
^{2}-\gamma_{2}\left\langle \left(  \hat{B}_{0}^{\dag}+\hat{B}_{0}\right)
^{2}\right\rangle \left\langle \hat{\jmath}_{3}\right\rangle \nonumber
\end{align}
Here
\begin{align*}
&  \varepsilon^{\prime}=-2\varepsilon+\gamma_{2}\left\langle \left(  \hat
{B}_{0}^{\dag}+\hat{B}_{0}\right)  ^{2}\right\rangle \\
&  \gamma_{2}^{\prime}=\gamma_{2}\left(  \left\langle \hat{\jmath}%
_{3}\right\rangle +\frac{1}{2}\right)
\end{align*}
This mean-field Hamiltonian, Eq. \ref{eq:MF}, differs from Eq.~34 in
Ref.~\onlinecite{Birman} and can be diagonalized using rotation, squeezing and
displacement transformations;%
\[
\hat{W}=e^{\alpha\left(  \hat{B}_{0}^{\dag}-\hat{B}_{0}\right)  }%
e^{\frac{\beta}{2}\left(  \hat{B}_{0}^{\dag2}-\hat{B}_{0}^{2}\right)
}e^{i\theta\hat{\jmath}_{1}}%
\]
These yield the following set of (self-consistent) coupled equations for the
parameters $\alpha,\beta$ and $\theta$%
\begin{equation}
\tan\theta=\frac{-4\Delta}{\varepsilon^{\prime}},\qquad\alpha=\frac
{-\gamma_{1}e^{\beta}}{\omega_{0}+\gamma_{2}^{\prime}},\qquad e^{4\beta}%
=\frac{\omega_{0}+\gamma_{2}^{\prime}}{\omega_{0}}.\label{eq:selfcons}%
\end{equation}
This set of equations can be reduced to a polynomial equation, which we solved
numerically. Using these numerical results we investigate the behavior of the
order parameters%
\begin{align}
\eta_{FE} &  \equiv\left\langle \hat{B}_{0}^{\dag}+\hat{B}_{0}\right\rangle
=2\alpha e^{-\beta}\\
\eta_{SC} &  \equiv\left\langle \hat{\jmath}_{2}\right\rangle =-\frac{1}%
{2}\sin\theta\nonumber
\end{align}
%

\begin{figure}
[ptb]
\begin{center}
\includegraphics[
height=5.1225cm,
width=6.1945cm
]%
{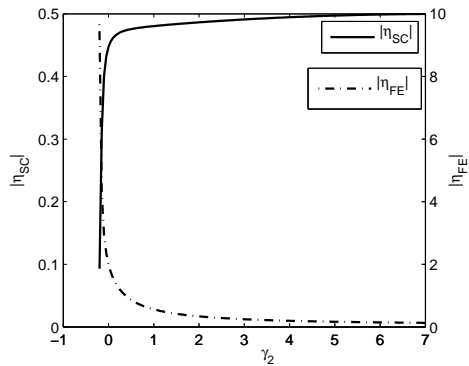}%
\caption{The absolute values of the superconducting order parameter and the
ferroelectric order parameter as a function of the coupling constant
$\gamma_{2}.$ The left side of the ordinate corresponds to small negative
values of $\gamma_{2}.$ Here $\Delta=\varepsilon=\gamma_{1}=\omega_{0}=1.$}%
\label{fig:orderpar}%
\end{center}
\end{figure}

Fig. \ref{fig:orderpar} shows the dependence of the two order parameters on
the coupling coefficient between the ferroelectric and superconducting
subsystems. There is a smooth evolution of each of the order parameters with
the coupling coefficient $\gamma_{2}.$ Both order parameters are non-vanishing
at $\gamma_{2}=0$. We note that $\gamma_{2}$ is bounded by two critical
values, beyond which there are no real solutions for the set of
self-consistent equations (\ref{eq:selfcons}). At the positive critical value
the superconducting gap parameter $\eta_{SC}$ reaches its maximum, and the
polarization order parameter $\eta_{FE}$ reaches its minimum. For negative
values of $\gamma_{2}$ the superconducting order parameter vanishes very
rapidly, while the ferroelectric order parameter sharply diverges, when
$\gamma_{2}$ approaches its negative critical value.

In conclusion we note that our new model for SC-FE coexistence/competition
does agree with the spirit of the Matthias conjecture that each of these
cooperative effects tends to exclude or suppress the other\cite{Matthias}.

\bigskip

\begin{acknowledgements}
JLB thanks the Institute of Theoretical Physics of Technion
for its support and hospitality. Also support from PSC-CUNY is acknowledged.
\end{acknowledgements}


\begin{thebibliography}{9}                                                                                                %


\bibitem {Birman}J.L. Birman and M. Weger, Phys. Rev. B \textbf{64}, 174503 (2001).

\bibitem {Matthias}B. Matthias, in \emph{Ferroelectricity}, edited by E.
Weller (Elsevier, New York, 1976)
\end{thebibliography}
\end{document}